\documentclass[aps,floatfix,twocolumn,prb,superscriptaddress]{revtex4-2}
\usepackage{xcolor} 
\usepackage{graphicx}
\usepackage[hidelinks]{hyperref}
\usepackage{amssymb}
\usepackage[utf8]{inputenc}
\usepackage{amsmath}
\usepackage{siunitx}
\usepackage{braket}
\usepackage{adjustbox, makecell}
\usepackage{dsfont}
\newcommand{\abs}[1]{\left|#1 \right|}

\renewcommand{\vec}[1]{\boldsymbol{#1}}
\newcommand{\iu}[0]{\mathrm{i}}

\newcommand{\NS}[1]{\textcolor{violet}{#1}}

\definecolor{vastkust}{RGB}{0, 48, 80} 
\hypersetup{
  colorlinks,
  citecolor=vastkust,
  linkcolor=vastkust,
  urlcolor=vastkust}

\begin{document}

\title{Unifying microscopic theories for the phono-magnetic effect}
\author{Natalia Shabala}
\affiliation{Department of Physics and Astronomy, Chalmers University of Technology, 412 96 G\"oteborg, Sweden}

\author{Finja Tietjen}
\affiliation{Department of Physics and Astronomy, Chalmers University of Technology, 412 96 G\"oteborg, Sweden}

\author{Ylva Liljegren}
\affiliation{Department of Physics and Astronomy, Chalmers University of Technology, 412 96 G\"oteborg, Sweden}

\author{R.~Matthias Geilhufe}
\affiliation{Department of Physics and Astronomy, Chalmers University of Technology, 412 96 G\"oteborg, Sweden}

\date{\today}

\begin{abstract}
Phonon angular momentum induces an effective magnetic field, a phenomenon called phono-magnetic effect, which has been measured to vary significantly in size.  Here, we compare three approaches for deriving the effective magnetic field of a phonon, using electron-phonon coupling and orbital magnetism. The adiabatic approach assumes a slow ionic motion, keeping electrons in the ground state. The perturbative approach treats the electron-phonon interaction as a perturbation to the electronic ground state, and the Floquet approach is based on the time-periodicity of the circular ionic motion. We show that all three approaches lead to the same effective Hamiltonian in the low-frequency limit, which moves us closer towards a unified theory of the phono-magnetic effect. 
Furthermore, we identify two phononic contributions to the sample magnetization, spontaneous and induced. Thus, we clarify the role of the effective magnetic field in the phonon-induced magnetization. We conclude by providing a numerical estimate for the effective magnetic field in SrTiO$_3$. 
\end{abstract}

\maketitle

\section{Introduction}
\begin{figure}
    \centering
    \includegraphics[width=\linewidth]{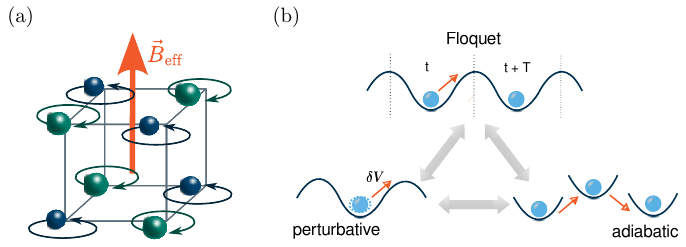}
    \caption{(a) Schematic illustration of the phono-magnetic effect. (b) All three approaches - Floquet, perturbative and adiabatic - are connected.}
    \label{fig:axial-phonons}
\end{figure}

Collective lattice excitations, i.e., phonons, refer to the thermal or driven ionic motion in solids. Phonons can possess a non-zero angular momentum \cite{McLellan1988AngularDynamics,Zhang2014AngularEffect}, called chiral or axial phonons~\cite{juraschek2025chiral,Zhang2015ChiralLattices}. Due to the ionic motion on circular orbits, axial phonons can carry a phonon magnetic moment~\cite{Schaack1975,rebane1983faraday,juraschek2017dynamical, Juraschek2019OrbitalPhonons,juraschek2020phono, Geilhufe2021DynamicallyKTaO_3} and induce an effective magnetic field~\cite{shabala2025axial,ren2021phonon,chaudhary2024giant,shabala2024phonon, Merlin2024MagnetophononicsMisnomer, Merlin2024UnravelingFields,klebl2024ultrafast}, termed \textit{phono-magnetic effect} (Fig. \ref{fig:axial-phonons} (a)). There is a wide range of experiments measuring the magnetization of systems with axial phonons, which cover different materials and set-ups.
The dynamical set-ups are based on laser-excited non-thermal axial phonons, where the emergent magnetization is measured with the magneto-optical Kerr-effect~\cite{Biggs2025UltrafastLiNbO3, Basini2024, Davies2024, luo2023large}.
In contrast, the static set-ups investigate thermal axial phonons by the phonon Zeeman effect, i.e., the splitting of degenerate phonon modes in the presence of a magnetic field~\cite{Schaack1975, Schaack1976, Schaack1977,juraschek2017dynamical,baydin2022magnetic, hernandez2023observation, lujan2024spin, mustafa2025origin, che2024magnetic, cheng2020large, wu2023fluctuation, wu2025magnetic}. 

To understand these observations, we follow the formalism on semiclassical dynamics developed in Ref.~\cite{ren2021phonon} and distinguish between two mechanisms for the formation of magnetization due to axial phonons. First, assuming charged ions, their circular motion induces a local magnetic moment, typically in the order of a nuclear magneton~\cite{juraschek2017dynamical, Juraschek2019OrbitalPhonons, Geilhufe2021DynamicallyKTaO_3}. Second, due to electron-phonon coupling, the phonon angular momentum gives rise to a splitting of orbital multiplets of electrons and an emergent orbital Zeeman effect in the presence of a pseudomagnetic field due to the phonon angular momentum. The dominance of either of the two contributions depends on the material and setup. In analogy to the terminology of magnets, we relate the first contribution to the \textit{spontaneous magnetization} and the second contribution to an \textit{induced magnetization}. The central scope of this paper is the derivation of the second contribution. 

In recent years, several specialized microscopic theories have been proposed and developed, predicting the splitting of electronic bands due to phonon angular momentum. 
These approaches are based on general assumptions on phonon dynamics (Fig.~\ref{fig:axial-phonons}(b)), such as perturbation theory assuming a small change of electronic states due to phonons \cite{chaudhary2024giant, fransson2023chiral, shabala2024phonon}, Floquet theory assuming the time-periodic dynamics of phonons \cite{klebl2024ultrafast}, or the adiabatic approach assuming that the timescales for phonon dynamics is much slower than the electron dynamics~\cite{ren2021phonon}. 
However, the connection between these different approaches has been missing, but would give a more coherent picture of the effect.

In this paper, we investigate the similarities between the perturbative, Floquet and adiabatic approaches, which we present in sections~\ref{sec:floquet} - \ref{sec:adiabatic}. 
As expected, the derived effective Hamiltonian, which captures the electronic energy splitting caused by electron-phonon coupling, is the same when the coupling is approximated as weak for the perturbative approach, and as time-periodic for Floquet (details given in section~\ref{sec:potential}).
In the limit of low phonon frequency, we obtain the result of the adiabatic approach, discussed in section~\ref{sec:adiabatic}. 
In section~\ref{sec:comparison}, we compare these results. In section~\ref{sec:magnetization} we relate the effective magnetic field due to the presence of an axial phonon to the semiclassical theory presented in Ref.~\cite{ren2021phonon} and derive its contribution to a sample magnetization. We close with an example, where we estimate the size of the effective magnetic field in laser-driven SrTiO$_3$, as discussed in section~\ref{sec:example}. This section also includes a discussion of the relevance of allowed orbital transitions on the effective splitting of the electronic levels. 

\section{Electron-Phonon Coupling and Notation}\label{sec:potential}
Phonons describe the collective excitation of the crystalline lattice. They are typically described as the small displacement $\vec{u}$ of an ion from its equilibrium position $\vec{R}_0$ (we neglect an index running over all ions for brevity). As a result, the potential describing the electrostatic interaction of ions and electrons is expanded in $\vec{u}$, 
\begin{equation}
    U(\vec{R}) \approx U(\vec{R}_0) +  \mathbf{u}(t) \cdot \nabla_{\mathbf{u}} U.
\end{equation}
Merging the bare potential $U(\vec{R}_0)$ with the kinetic energy of the electrons into a Hamiltonian $H_0$, the resulting Schrödinger equation is
\begin{equation}\label{seq}
    \left[H_0 + V(t)\right]\ket{\Psi(t)} = \mathrm{i}\hbar \frac{\partial}{\partial t} \ket{\Psi(t)},
\end{equation}
where we have introduced the time-dependent perturbation
\begin{equation}\label{time}
     V(t) = \mathbf{u}(t) \cdot \nabla_{\mathbf{u}} U.
\end{equation}
The time-dependent ionic displacement characterizing axial phonons is written in complex form, 
\begin{equation}\label{eq:u_circ_pol}
    \vec{u} = \underbrace{\left(\frac{1}{\sqrt{2}}u_R (\vec{\hat{e}_x} + i\vec{\hat{e}}_y) + \frac{1}{\sqrt{2}}u_L (\vec{\hat{e}_x} - i\vec{\hat{e}}_y)\right)}_{\vec{u_0}}e^{i\omega t}.
\end{equation}
Here, $\omega$ is the phonon frequency and $u_{R,L} = \frac{1}{\sqrt{2}}(u_x \mp \iu u_{y})$ is the amplitude of the right(left)-circularly polarized mode, while $x$ and $y$ label two degenerate and orthogonal phonon modes. 
Throughout the paper, we consider the special case 
\begin{equation}\label{eq:axialphon}
    \vec{u}_{R,L} = u(\cos{(\omega t)}, \pm\sin{(\omega t)}, 0)^\text{T},
\end{equation}
describing a right- or left-circularly polarized coherent phonon mode, where $u$ is the total amplitude.
Hence, using the explicit form of the displacement for a circularly polarized and periodically evolving phonon displacement Eq.~\eqref{eq:axialphon}, the time-dependent perturbation becomes
\begin{equation}\label{eq:potential-periodic}
    V(t) = u \cos{(\omega t)} \, \partial_{u_x} U \pm u \sin{(\omega t)} \, \partial_{u_y} U.
\end{equation}
It is convenient to express the sine and cosine functions in terms of their complex representation, leading to
\begin{equation}\label{eq:potential-periodic:complex}
    V(t) = v e^{\iu \omega t} + v^* e^{-\iu \omega t}, \quad v = \frac{u}{2}\left[\partial_{u_x}U \mp \iu \partial_{u_y}U\right].
\end{equation}
Finally, we note that we subsequently assume $\vec{u}$ to be in units of a phonon mode, i.e., $\left[\vec{u}\right] = \text{\AA} \sqrt{\text{a.m.u.}}$, with a.m.u. denoting the atomic mass unit. This allows us to define the ionic angular momentum (phonon angular momentum) as follows,
\begin{equation}
    \vec{L} = \vec{u}\times\dot{\vec{u}}.
\end{equation}

\section{Three approaches for computing the effective magnetic field induced by axial phonons}

In the following section, we derive the effective electronic Hamiltonian including electron-phonon interaction.
While these approaches were originally proposed in previous work~\cite{klebl2024ultrafast, shabala2024phonon, xiao2010berry}, here we show that the three methods lead to the same expression of the effective Hamiltonian, when the same form of electron-phonon coupling, discussed in section~\ref{sec:potential}, is assumed.

\subsection{Floquet perturbation theory}\label{sec:floquet}
Floquet theory offers a solution of the Schrödinger equation for a time-periodic Hamiltonian $H(t+T) = H(t)$~\cite{oka2019floquet}. Here, we assume the coherent, circularly polarized phonon mode with frequency $\omega$, similar to Ref.~\cite{klebl2024ultrafast}.
According to Floquet's theorem, the solution to the Schrödinger equation Eq.~\eqref{seq} is given by a product of a phase factor and a time-periodic function,
\begin{equation}\label{eq:floquet}
    \ket{\psi_\alpha(t)} = e^{- i \varepsilon_\alpha t} \ket{\phi_\alpha(t)},
\end{equation}
where $\ket{\phi_\alpha(t)} = \ket{\phi_\alpha(t+T)}$ is called the Floquet mode for a specific mode $\alpha$, and $\varepsilon_\alpha$ is a conserved quantity known as the quasienergy. Using Eq.~\eqref{eq:floquet} in the Schrödinger equation Eq.~\eqref{seq}, the Floquet modes are shown to satisfy the Floquet-Schrödinger equation
\begin{equation}
    [H(t) - i \hbar \partial_t] \ket{\phi_\alpha(t)} = \varepsilon_\alpha \ket{\phi_\alpha(t)}.
\end{equation}
Because the Floquet modes are periodic in time, they are expanded in a Fourier series $\ket{\phi_\alpha(t)} = \sum_n e^{in\omega t} \ket{\phi_\alpha^n}$. Inserting this expansion into the Floquet Schrödinger equation gives, for each mode $\alpha$,
\begin{equation}
    \sum_n e^{i n \omega t} [H(t) + \hbar n \omega] \ket{\phi_\alpha^n}
    = \varepsilon_\alpha \sum_n e^{i n \omega t} \ket{\phi_\alpha^n}.
\end{equation}

Multiplying both sides by $e^{-i m \omega t}$ and integrating over one period $T$, we obtain 
\begin{equation} \label{eq:fourier_harmonics}
    \sum_n \left( h_{m-n} + \hbar m \omega \, \delta_{mn} \right) \ket{\phi^n_\alpha}
    = \varepsilon_\alpha \ket{\phi^m_\alpha},
\end{equation}
where we define the Floquet hoppings
\begin{equation}\label{eq:hoppings}
     h_{m-n} \equiv \frac{1}{T} \int_0^T H(t) e^{-i (m-n) \omega t} \, dt.
\end{equation}
From the time-dependent Hamiltonian in the Schrödinger equation Eq.~\eqref{seq} and the complex representation of the time-dependent perturbation given in Eq.~\eqref{eq:potential-periodic:complex}, we find that the only non-zero Fourier coefficients in the Floquet hoppings Eq.~\eqref{eq:hoppings} occur for $m-n = 0, \pm 1$:
\begin{equation} \label{eq:fourier_coeffs}
\begin{cases}
     h_0= H_0, \\[4pt]
    h_{-1} = v \equiv h_1 \\[4pt]
    h_{+1} = v^* \equiv h_1^\dagger.
\end{cases}
\end{equation}
Substituting into Eq.~\eqref{eq:fourier_harmonics} gives the analytic recursion relation
\begin{equation} \label{eq:recursion}
    (h_0 + \hbar m \omega) \ket{\phi_\alpha^m} + h_1^\dagger \ket{\phi_\alpha^{m-1}} + h_1 \ket{\phi_\alpha^{m+1}}
    = \varepsilon_\alpha \ket{\phi_\alpha^m},
\end{equation}
which connects all Fourier components $\phi_\alpha^m$ of a given Floquet mode $\alpha$. Because the recursion formula Eq.~\eqref{eq:recursion} couples only ``nearest neighbors'', the Floquet Hamiltonian is tridiagonal,
\begin{equation}
\resizebox{0.85\linewidth}{!}{$
\mathcal{H}_F =
\begin{pmatrix}
\ddots & \ddots &        &        &        &        \\
\ddots & h_0-2\hbar\omega & h_1 &        &        &        \\
       & h_1^\dagger & h_0 -\hbar\omega & h_{1} &        &        \\
       &        & h_1^\dagger & h_0 & h_1 &        \\
       &        &        & h_1^\dagger & h_0 + \hbar\omega & h_1 \\
       &        &        &        & h_1^\dagger & h_0 +2\hbar\omega & \ddots \\
       &        &        &        &        & \ddots & \ddots
\end{pmatrix}
$} .
\end{equation}
We are primarily interested in the zeroth component $\phi_\alpha^0$, which corresponds to perturbations of the ground state. 
To this end, we construct an effective Hamiltonian acting only on $\phi_\alpha^0$, while incorporating the effects of the neighboring Fourier components. 

We start from the $m=0$ case of Eq.~\eqref{eq:recursion} and recursively eliminate the higher-order Fourier components $\phi_\alpha^{\pm 1}, \phi_\alpha^{\pm 2}, \ldots$ by solving the corresponding $m=\pm 1, \pm2 , \ldots$ equations and substituting the results back. 
Solving the $m=\pm 1$ equations then yields continued-fraction self-energies of the form 
\begin{equation}
    \ket{\phi_\alpha^1} = G_{-}(\varepsilon_\alpha) h^\dagger_1 \ket{\phi_\alpha^0}, \quad \ket{\phi_\alpha^{-1}} = G_{+}(\varepsilon_\alpha) h_1 \ket{\phi_\alpha^0},
\end{equation}
where $G_{\pm}(\varepsilon_\alpha)$ are given by
\begin{equation} \label{eq:Greens}
    \begin{aligned}
            G_{+}(\varepsilon_\alpha) &= \frac{1}{\varepsilon_\alpha - h_0 + \hbar \omega - h^\dagger_{1} \frac{1}{\varepsilon_\alpha - h_0 + 2 \hbar \omega - \ldots } h_{1}} \\
         G_{-}(\varepsilon_\alpha) &= \frac{1}{\varepsilon_\alpha - h_0 - \hbar \omega - h_{1} \frac{1}{\varepsilon_\alpha - h_0 - 2 \hbar \omega - \ldots } h^\dagger_{1}} 
    \end{aligned}
\end{equation}
Inserting $\phi_\alpha^{\pm 1}$ into the $m=0$ equation in Eq.~\eqref{eq:recursion} gives
\begin{equation} \label{eq:Heff}
 H_\text{eff} = h_0 + h_1 G_{+}(\varepsilon_\alpha) h_1^\dagger + h_1^\dagger G_{-}(\varepsilon_\alpha) h_1.
\end{equation}
Since the Floquet hoppings are linear in the phonon displacement, $h_{1} \propto u$, and the displacements are small, contributions from higher sidebands ($|m| \geq 2$) involve additional powers of $u$ and therefore become progressively smaller. Thus, we truncate the continued fractions at the first sidebands to obtain a leading-order effective Hamiltonian
\begin{equation}
     H_\text{eff} \simeq
    h_0 + h_1 \frac{1}{\varepsilon_\alpha - h_0 - \hbar \omega} h_1^\dagger
    + h_1^\dagger \frac{1}{\varepsilon_\alpha - h_0 + \hbar \omega} h_1.
\end{equation}
To evaluate this Hamiltonian, we consider a generic orbital basis $\{\ket{a}, \ket{b}, \ldots\}$, in which $h_0$ is diagonal, $\bra{n} h_0 \ket{m} = E_n \delta_{nm}$. Corrections to the Floquet quasienergies appear at second order in the phonon displacement, such that $\varepsilon_\alpha = E_\text{initial} + \mathcal{O}(u^2)$. Since the effective Hamiltonian is already second order in $h_1, h_1^\dagger$, we consistently replace $\varepsilon_\alpha$ by the unperturbed orbital energy of the reference state, $E_b$. The matrix elements in this orbital basis are then
\begin{equation}
\begin{aligned}
    \langle a | H_\text{eff} | b \rangle
&= \langle a | h_0 | b \rangle \\
& \quad 
+ \sum_{m,n} 
\langle a | h_1 | n \rangle
\left[ \frac{1}{E_b - h_0 - \hbar \omega} \right]_{nm} 
\langle m | h_1^\dagger | b \rangle \\
&\quad 
+ \sum_{m,n}
\langle a | h_1^\dagger | n \rangle
\left[ \frac{1}{E_b - h_0 + \hbar \omega} \right]_{nm} 
\langle m | h_1 | b \rangle.
\end{aligned}
\end{equation}
Here, we use the notation
\begin{equation}
    \left[ \frac{1}{(E_b - E_n) + \hbar \omega} \right] \delta_{nm} = \left[ \frac{1}{E_b - h_0 + \hbar \omega} \right]_{nm} \,.
\end{equation}
Now, since $h_0$ is diagonal, the double sums collapse to single sums. The first term only shifts the on-site energies and does not contribute to orbital energy splittings in the phono-magnetic effect, leaving the relevant part of the Hamiltonian as
\begin{equation}\label{eq:fourier:general}
\langle a | H_\text{eff} | b \rangle
= \sum_n \left[\frac{\langle a| h_1 | n \rangle \langle n | h_1^\dagger | b \rangle}{(E_b -E_n) - \hbar \omega}+ \frac{\langle a| h_1^\dagger | n \rangle \langle n | h_1 | b \rangle}{(E_b -E_n) + \hbar \omega}\right].
\end{equation}
Next, we insert the explicit forms of \(h_1\) and \(h_1^\dagger\) from Eq.~\eqref{eq:fourier_coeffs}. For convenience, we define the two numerators as  
\[
X = \langle a| h_1 | n \rangle \langle n | h_1^\dagger | b \rangle, \quad
Y = \langle a| h_1^\dagger | n \rangle \langle n | h_1 | b \rangle,
\]  
and denote \(E_{bn} = E_b - E_n\). Then each term can be rewritten as  
\begin{equation}
\begin{aligned}
    &\frac{X}{E_{bn} - \hbar \omega} + \frac{Y}{E_{bn}+ \hbar \omega} \\
= &\frac{E_{bn}}{E_{bn}^2 - \hbar^2 \omega^2} (X+Y)
+ \frac{\hbar \omega}{E_{bn}^2 - \hbar^2 \omega^2} (X-Y).
\end{aligned}
\end{equation}
The second term is antisymmetric under exchange of the spatial displacement components $u_x \leftrightarrow u_y$, which follows from the structure of $h_1$ in Eq.~\eqref{eq:fourier_coeffs} when evaluating the difference $X-Y$. 
This antisymmetric contribution therefore corresponds to the time-reversal-odd part. Since the magnetization breaks time-reversal symmetry, we focus on this term in the following. 

Noting that $Y = X^*$, we have a purely imaginary contribution $X-Y = 2 i \, \text{Im}(X)$, which leads to the final expression for $\langle a | H_\text{eff} | b \rangle \equiv H_\text{eff}^{ab}$ as
\begin{multline}\label{eq:floquet_final}
 H_\text{eff}^{ab}
= \frac{i \hbar}{2}L_z\, 
\sum_n \Bigg[
\frac{
\langle a|\partial_{u_x}U|n\rangle \langle n|\partial_{u_y}U|b\rangle
}{
E_{nb}^2 - \hbar^2 \omega^2
} \\
- \frac{
\langle a|\partial_{u_y}U|n\rangle \langle n|\partial_{u_x}U|b\rangle}{E_{nb}^2 - \hbar^2 \omega^2}  \Bigg].
\end{multline}
Here, we have used the Floquet coefficients as defined by Eq.~\eqref{eq:fourier_coeffs}, with the perturbation amplitude given by Eq.~\eqref{eq:potential-periodic:complex}.
Since this effective Hamiltonian is derived from the Hermitian Floquet Hamiltonian, it is Hermitian by construction. 

\subsection{Time-dependent perturbation theory}\label{sec:perturbation}
We continue with the perturbative approach, assuming that the time-dependent perturbation $V(t)$ is sufficiently weak. To do so, we transform to the interaction picture, $\overline{\Psi}(t) = e^{\frac{\iu}{\hbar}H_0t}\Psi(t)$ and $\overline{V}(t) = e^{\frac{\iu}{\hbar}H_0t}V(t)e^{-\frac{\iu}{\hbar}H_0t}$ with the time-evolution of the state given by
\begin{equation}
    \iu \hbar \frac{\partial \overline{\Psi}(t)}{\partial t} = \overline{V}(t)\overline{\Psi}(t).
\end{equation}
We obtain the solution iteratively, with all terms up to second order being \cite{pershan1966theoretical,popova2014microscopic, Wong2025}
\begin{equation}
    \overline{\Psi}(t) = \left[1 - \frac{\iu}{\hbar} \int_0^t dt' \overline{V} - \frac{1}{\hbar^2}\int_{0}^t dt' \overline{V} \int_0^t dt''\overline{V}\right]\overline{\Psi}(0).
\end{equation}
 This expansion allows us to define an effective Hamiltonian in terms of matrix elements between electronic states $\ket{a}$ and $\ket{b}$, 
\begin{multline}
     \bra{a} \left( 1-\frac{i}{\hbar}\int_0^t\mathrm{d}t'\,\overline{H}_\text{eff} (t) \right) \ket{b} \\ =\bra{a} \left( 1-\frac{\iu}{\hbar} \int_0^t dt'\overline{V}(t') \qquad\qquad\right. \\ - \left. \frac{1}{\hbar^2} \int_0^tdt' \overline{V}(t') \int_0^t dt''\overline{V}(t'') \right) \ket{b}.
\end{multline}
As the phonon frequency is typically not resonant with an electronic transition, transition processes in first order are neglected. Hence, the relevant term is the contribution to second order, leading to the explicit form of the effective Hamiltonian,
\begin{equation}
     \bra{a} \overline{H}_\text{eff} (t) \ket{b} =\frac{\mathrm{i}}{\hbar}\sum_n \bra{a} \overline{V} \ket{n} \int_0^t dt'\bra{n}\overline{V}(t') \ket{b}.
\end{equation}
Transforming back to the Schrödinger picture gives
\begin{multline}
     \bra{a} H_\text{eff} (t) \ket{b} =\frac{\mathrm{i}}{\hbar}\sum_n  \bra{a} V \ket{n} e^{\iu \omega_{bn}t}\,\\\times \int_0^t dt' e^{\iu \omega_{nb}t'} \bra{n}V(t') \ket{b},
\end{multline}
where we use the notation $H_0 \ket{x} = \hbar \omega_x \ket{x}$ and $\omega_{xy}=\omega_x - \omega_y$. Using the explicit form of $V$ from Eq.~\eqref{eq:potential-periodic:complex}, we perform the time integral. 
To do so, we assume that $v(t)$ is varying slowly compared to $\abs{\omega_{bn}\pm\omega}$ and that $v$ is switched on adiabatically, $v(-\infty) = 0$. Also, we neglect terms corresponding to second harmonic generation $\sim e^{\pm\iu 2\omega t}$ and obtain
\begin{equation} \label{eq:Heff_general_ife}
\begin{split}
    \bra{a}H_{\text{eff}}(t)\ket{b}  = \sum_n &\left[\frac{\bra{a}v\ket{n}\bra{n}v^*\ket{b}}{E_{nb}-\hbar\omega}\right. \\
    & + \left.\frac{\bra{a}v^*\ket{n}\bra{n}v\ket{b}}{E_{nb}+\hbar\omega}\right].
\end{split}
\end{equation}
Considering the Fourier coefficients defined in Eq.~\eqref{eq:fourier_coeffs}, we find that Eq.~\eqref{eq:Heff_general_ife} is identical to Eq.~\eqref {eq:fourier:general}. 
Hence, following the same antisymmetrization scheme to find the time-reversal symmetry-breaking solution as presented in the previous section, we end up with the effective Hamiltonian
\begin{multline}\label{eq:H_eff_IFE}
 H_\text{eff}^{ab}
= \frac{i \hbar}{2}L_z\, 
\sum_n \Bigg[
\frac{
\langle a|\partial_{u_x}U|n\rangle \langle n|\partial_{u_y}U|b\rangle
}{
E_{nb}^2 - \hbar^2 \omega^2
} \\
- \frac{
\langle a|\partial_{u_y}U|n\rangle \langle n|\partial_{u_x}U|b\rangle}{E_{nb}^2 - \hbar^2 \omega^2}  \Bigg].
\end{multline}

\subsection{Adiabatic evolution of the electronic states}\label{sec:adiabatic}
In an adiabatic process, the system parameters change slowly enough to remain in the same eigenstate throughout the time evolution \cite{xiao2010berry} (lower right of Fig.~\ref{fig:axial-phonons}(b)). The typical phonon frequency is on the order of \si{\tera\hertz}, corresponding to an ionic motion on a timescale of \si{\pico\second}. 
In contrast, a typical electron energy level is on the scale of \si{\electronvolt} with a timescale of \si{\femto \second} for the evolution of electronic states, thus, the change in the electronic state induced by the interaction with the phonon can be considered sufficiently slow compared to the typical timescale of an electron.
    
The adiabatic approach has been used to derive the magnetization induced by axial phonons through studying the relationship between the current and the orbital magnetization \cite{ren2021phonon, king-smith1993theory, yao2025theory}.
It is possible to show that during the adiabatic time evolution of the electrons in the coupling potential, the electronic states gain a geometric phase that is proportional to the phonon angular momentum \cite{shabala2025axial}. 
In this section, we derive this geometric phase and show that it is related to the results of sections \ref{sec:floquet} and \ref{sec:perturbation}.

To examine the time evolution of the electronic states, we turn to the Schrödinger equation: 
\begin{equation}
    H_{\vec{u}(t)}\ket{\Psi} = i\hbar \partial_t \ket{\Psi},
\end{equation}
where the ion displacement $\vec{u}$ is now a parameter of the Hamiltonian.
In a generic crystal, the energy levels are allowed to be degenerate, i.e., $H(t)\ket{n^{g_n}(t)} = E_n(t)\ket{n^{g_n}(t)}$, with $g_n$ running over degenerate states with energy $E_n$.  
Therefore, the solutions of the Schrödinger equation take the form \cite{rigolin2010adiabatic, rigolin2012adiabatic}:
\begin{equation}\label{eq:berry_psi}
    \ket{\Psi(t)} = \sum_n \sum_{g_n} e^{-i\omega_n (t)} b_n(0) \text{W}^n_{h_n g_n} \ket{n^{g_n}(t)}.
\end{equation}
Here, $\omega_n(t) = \frac{1}{\hbar}\int_0^tE_n(t')dt'$ is the dynamical phase and $\text{W}^n_{h_n g_n}$ are elements of a unitary matrix $\vec{W}^n$, such that
\begin{equation}\label{eq:Un}
    \vec{W}^n(t) = \vec{W}^n(0) \mathcal{T}\exp{\left(\vec{A}^{nn}(t')dt'\right)}.
\end{equation}
Here, $\mathcal{T}$ is the time-ordering operator and $\vec{A}^{nn}$ is a matrix with elements
\begin{equation}\label{eq:A_mn}
    A_{h_n g_m}^{mn}(t) = - \bra{m^{g_m}(t)} \partial_t \ket{n^{h_n}(t)}. 
\end{equation}
Using Eqs.~\eqref{eq:A_mn} and \eqref{eq:Un} in Eq.~\eqref{eq:berry_psi}, allows us to write the time-evolved electronic states as:
\begin{equation}
\begin{split}
    \ket{\Psi(t)} & = \sum_{g_n} \sum_n e^{-i\omega_n(t)}b_n(0) \\
    &  \times \mathcal{T} \exp{\left(-\int_{0}^t \bra{g_n}\partial_{t'} \ket{h_n}dt'\right)}\ket{n^{g_n}(t)}.
\end{split}
\end{equation}
We assume that the system is initialized in the ground state multiplet, which allows us to set $b_n(0) = 1$ \cite{rigolin2010adiabatic, rigolin2012adiabatic}.

The evolution of the electronic states is adiabatic. Therefore, the state is confined to the ground state multiplet, $A^{mn}_{g_m h_n} (t) = \delta_{mn} A^{nn}_{g_n h_n} (t)$ \cite{rigolin2010adiabatic}. This allows us to relabel the indices as $g_n \rightarrow a$, $h_n \rightarrow b$ for convenience. Note that the corresponding states $\ket{a}$, $\ket{b}$ depend on the displacement $\vec{u}$ in general. With the initial state being $\ket{\psi_0}$, this gives
\begin{multline}\label{eq:psi_omega_gamma}
    \ket{\Psi(t)}  = \mathcal{T} \sum_m \exp{\left(-\iu\omega_0t\right)}\\
     \times \exp{\left(-\int_0^t\bra{a}\partial_{t'}\ket{b}dt'\right)}\ket{\psi_0} \\
    = \mathcal{T} \sum_m \exp{\left(-\iu\omega_0t\right)}\exp{\left(-\iu\gamma_{ab}(t)\right)}\ket{\psi_0}.
\end{multline}
In Eq.~\eqref{eq:psi_omega_gamma} we identified the non-Abelian geometric phase, given by
\begin{equation}
    \gamma_{ab}(t) = -\text{i}\int_0^t \bra{a}\partial_{t'} \ket{b}dt'.
\end{equation}
Note that $\vec{\gamma}$ is a Hermitian matrix because $\bra{a}\partial_{t'} \ket{b}$ are elements of an anti-Hermitian matrix. Consequently, $\vec{\gamma}$ has real eigenvalues. It follows that it can always be transformed into a diagonal matrix with real elements, corresponding to the acquired geometric phase for each state.

Now we examine the contribution of axial phonons to this phase. 
To do that, we first write the time derivative in terms of the displacement $\vec{u}$
\begin{equation}
    \frac{\partial}{\partial t'} = \frac{\partial}{\partial \vec{u}}\frac{\partial \vec{u}}{\partial t'}.
\end{equation}
Hence, the geometric phase now takes the form
\begin{equation}
    \gamma_{ab}(t) =-\text{i} \int_0^t \mathrm{d}t \bra{a}\nabla_{\vec{u}} \ket{b} \dot{\vec{u}}.
\end{equation}
As before, it is reasonable to assume a small displacement, which allows us to use the Taylor expansion in $\vec{u}$ up to the first order
\begin{equation}
    \gamma_{ab} \approx \gamma_{ab}^0 + \sum_{j} u_j\,\partial_{u_j}\gamma_{ab}.
\end{equation}
Assuming that at zero displacement there is no geometric phase, we can neglect the first term and write $\gamma_{mn}(t)$ as:
\begin{equation}
    \gamma_{ab}(t) \approx -\text{i} \sum_{ij} \int_0^t \mathrm{d}t\, \left(\partial_{u_j} A^{ab}_{u_i}\right)\, \dot{u}_i u_j,
\end{equation}
where $A^{ab}_{u_i} = \bra{a}\partial_{u_i} \ket{b}$ is the non-Abelian phonon Berry connection. We now express the non-Abelian geometric phase as a sum of its symmetric and antisymmetric parts, given by
\begin{multline}
    \gamma_{ab}(t) = -\text{i} \int_0^t \mathrm{d}t\\ \times \sum_{ij} \left[ \frac{1}{2} \left\{\left(\partial_{u_j} A^{ab}_{u_i}\right)\,\dot{u}_i u_j + \left(\partial_{u_i} A^{ab}_{u_j}\right)\, \dot{u}_i u_j\right\} \right.
    \\\left.+ \frac{1}{2}\left\{\left(\partial_{u_j} A^{ab}_{u_i}\right)\,\dot{u}_i u_j - \left(\partial_{u_i} A^{ab}_{u_j}\right)\,\dot{u}_i u_j\right\}  \right].
\end{multline}
Consistent with the discussion in section \ref{sec:floquet}, we focus only on the antisymmetric part since it corresponds to the time-reversal-odd contribution. 
Additionally, we assume an ionic motion in $xy$-plane, similar to Eq.~\eqref{eq:axialphon}, 
\begin{multline}
    \gamma_{ab}(t) = -\frac{\iu}{2} \int_0^t dt' (u_x \dot{u}_y - \dot{u}_x u_y)  \\
      \times \left( \partial_{u_x} \bra{a}\partial_{u_y} \ket{b} - \partial_{u_y} \bra{a}\partial_{u_x} \ket{b} \right).
\end{multline}
We express this equation in terms of the Hamiltonian by writing \cite{xiao2010berry}
\begin{equation}
    \braket{\nabla_{\vec{u}}n|n'} = \frac{\bra{n}\nabla_{\vec{u}}H\ket{n'}}{(\varepsilon_n - \varepsilon_{n'})}.
\end{equation}
This gives us the final expression for the geometric phase
\begin{equation}\label{eq:adiabatic_final}
    \gamma_{ab}(t) = \NS{-\frac{1}{2}} \int_0^t dt' (u_x \dot{u}_y - \dot{u}_x u_y) \Omega_{u_x u_y}^{ab},
\end{equation}
where we have introduced the non-Abelian phonon Berry curvature
\begin{multline}\label{eq:BerryCurv}
    \Omega_{u_xu_y}^{ab}  = \iu \sum_{n\neq a}\left[\frac{\bra{a}\partial_{u_x}U \ket{n}\bra{n}\partial_{u_y}U \ket{b}}{{E_{nb}^2}}\right.\\
     - \left.\frac{\bra{a}\partial_{u_y}U \ket{n}\bra{n}\partial_{u_x}U \ket{b}}{E_{nb}^2}\right].
\end{multline}

Eq.~\eqref{eq:adiabatic_final} is related to the effective Hamiltonian derived in Eqs.~\eqref{eq:floquet_final} and \eqref{eq:H_eff_IFE}. To show this, we consider the time-evolution of the state $\ket{\Psi (t)}$, given by
\begin{equation}\label{eq:psi_phase_H}
    \ket{\Psi(t)} = \mathcal{T} \exp{\left(-\frac{\iu}{\hbar}\int_0^t dt'H(t')\right)} \ket{\Psi(0)}.
\end{equation}
By comparing with the equation for the time-evolved state given by Eq.~\eqref{eq:psi_omega_gamma}, we express the integral as 
\begin{equation}\label{eq:phase_H}
    \frac{1}{\hbar}\int_0^t dt'H(t') =  \frac{1}{\hbar}\int_0^tdt'H_0 +  \frac{1}{\hbar}\int_0^tdt'H_\text{eff}.
\end{equation}
Here, the first term is the dynamical phase, $ \frac{1}{\hbar}\int_0^tdt'H_0 =  \omega_0 t$, while the second term is the geometric phase of Eq.~\eqref{eq:adiabatic_final}. 
Thus, we obtain the effective Hamiltonian
\begin{multline}\label{eq:H_eff_adiabatic}
    H_\text{eff}  = \frac{\iu \hbar}{2} L_z \sum_{n'\neq m}\left[\frac{\bra{a}\partial_{u_x}U \ket{n}\bra{n}\partial_{u_y}U \ket{b}}{{E_{nb}^2}}\right.\\
     - \left.\frac{\bra{a}\partial_{u_y}U \ket{n}\bra{n}\partial_{u_x}U \ket{b}}{E_{nb}^2}\right],
\end{multline}
where $L_z = u_x \dot{u}_y - \dot{u}_x u_y$ is the phonon angular momentum. We note that Eq.~\eqref{eq:H_eff_adiabatic} is identical to Eqs.~\eqref{eq:floquet_final} and \eqref{eq:H_eff_IFE} in the low frequency regime, $\hbar\omega \ll \abs{E_{nb}}$. Finally, with the phonon angular momentum and the Berry curvature, we can write down the condensed form of the effective Hamiltonian given by
\begin{equation}\label{eq:H_L_Berry_curvature}
    H_\text{eff}^{ab} = \frac{\hbar}{2}L_z \Omega_{u_xu_y}^{ab},
\end{equation}
where $a$ and $b$ label the states in the degenerate subspace of the ground state multiplet.

\subsection{Comparison of the approaches}\label{sec:comparison}

In this section, we discuss the relationship between the Floquet, adiabatic and perturbative approaches to further highlight the agreement between them.
First, we note that all three formalisms examine the effect of the axial phonons on the electronic energy levels by deriving an expression for the effective Hamiltonian.
Additionally, as stated in section \ref{sec:potential}, we have assumed the same form of electron-phonon coupling as the starting point.
However, the three approaches rely on different assumptions about the change induced by the phonons. As illustrated in Fig. \ref{fig:axial-phonons}, the perturbative approach is based on the assumption that since the ion displacement characterizing the phonons is small, the change in the Hamiltonian is also small. 
The Floquet approach utilizes the fact that the presence of axial phonons makes the Hamiltonian periodic in time.
Finally, the underlying assumption of the adiabatic approach is that the induced change can be considered to be very slow.
Despite these different assumptions, in the low-frequency regime, the effective Hamiltonian derived using the different formalisms reduces to the same expression
\begin{multline}\label{eq:H_eff_comparison}
        H_{\text{eff}}^{ab}  =  \frac{\iu \hbar}{2} L_z 
    \sum_n \left[\frac{\bra{a}\partial_{u_x} U\ket{n}\bra{n}\partial_{u_y}U \ket{b}}{E_{nb}^2-\hbar^2\omega^2}\right. \\ - \left. \frac{\bra{a}\partial_{u_y} U\ket{n}\bra{n}\partial_{u_x} U \ket{b}}{E_{nb}^2-\hbar^2\omega^2} \right].
\end{multline}
As discussed earlier, the energy scale of a phonon is often much smaller than the energy scale of an electron, and therefore $E_{nb}^2 \gg \hbar^2 \omega^2$, which makes the low-frequency assumption reasonable. It is also worth noting that the influence of the phonon angular momentum on the electronic levels decreases quadratically with the increasing electronic band gap, although inverse cubic dependence on the gap has also been recently suggested~\cite{urazhdin2025atomic}. Additionally, $H_\text{eff}$ is proportional to the phonon angular momentum, $L_z$, and to the square of the magnitude of the electron-phonon coupling strength $g_{an\nu} \sim \bra{a}\partial_{u_\nu}\ket{n}$. In the opposite case of vanishing gap, e.g., in Dirac materials, the frequency term $\hbar^2\omega^2$ plays an important role in avoiding the singularity in the effective Hamiltonian. The potential divergence is discussed in, e.g., Ref. \cite{ren2021phonon}.

\section{The effective magnetic field and magnetization}\label{sec:magnetization}

In the following, we relate the effective Hamiltonian Eq.~\eqref{eq:H_eff_comparison} to the magnetization due to the second topological current as derived by Ren \textit{et al.}~\cite{ren2021phonon}. 
Using semi-classical dynamics, the total magnetization is~\cite {ren2021phonon}
\begin{equation}
   M _z = \frac{e}{2}L_z \int \frac{d\vec{k}}{(2\pi)^2}\Omega_{k_xk_yu_xu_y}.
\end{equation}
Here, $\Omega_{k_xk_yu_xu_y}$ is the second Chern form, defined as $\Omega_{k_xk_yu_xu_y} = \Omega_{k_xu_y}\Omega_{k_yu_x} - \Omega_{k_xu_x}\Omega_{k_yu_y} + \Omega_{k_xk_y}\Omega_{u_xu_y}$ \cite{ren2021phonon}. We use the notation $\Omega_{mn} = \iu\left(\partial_m \bra{\Psi}\partial_n\ket{\Psi} - \partial_n \bra{\Psi}\partial_m\ket{\Psi}\right)$. 
The magnetization consists of two terms
\begin{equation}\label{mag:total}
    M_z = M_z^{(1)} + M_z^{(2)},
\end{equation}
where
\begin{equation}\label{eq:M_orbital}
    M_z^{(1)} = \frac{e}{2}L_z \int_{\text{BZ}} \frac{d\vec{k}}{(2\pi)^2} \left[\Omega_{k_xu_y}\Omega_{k_yu_x} - \Omega_{k_xu_x}\Omega_{k_yu_y}\right],
\end{equation}
contains all Berry curvatures with mixed contributions in $\vec{k}$ and $\vec{u}$, and 
\begin{equation}\label{eq:M_topological}
    M_z^{(2)} = \frac{e}{2}L_z\int\frac{d\vec{k}}{(2\pi)^2}\Omega_{k_xk_y}\Omega_{u_xu_y}
\end{equation}
contains the non-mixed contributions.
In the following paragraphs, we identify these two terms of the magnetization as spontaneous and induced.

\subsection{Spontaneous magnetization of axial phonons}
To understand the first term $M_z^{(1)}$, we note that the polarization of a material can be written in terms of the Berry connection~\cite{king-smith1993theory, Resta1994ModernFerroelectrics,xiao2010berry}, $\vec{P} = -e \int_{\text{BZ}} \mathrm{d}\vec{k}\,\bra{\psi_{\vec{k}}} \iu \nabla_{\vec{k}}\ket{\psi_{\vec{k}}} |_{\lambda(0)}^{\lambda(T)}$, where $\lambda(t)$ is the parameter of the state. 
In our case, this parameter is the ionic displacement. 
The Born effective charge is the response of the polarization under ionic displacement. Under periodic boundary conditions, it can be written in terms of the mixed Berry connection $\Omega_{k_j u_i}$,
\begin{equation}
    Z_{ij} = V_0\,\frac{\partial P_j}{\partial u_{i}} = e V_0 \int \frac{d\vec{k}}{(2\pi)^2} \Omega_{k_j u_i}.
\end{equation} 
The latter form allows us to introduce a Born effective charge density, given by
\begin{equation}
    Z_{ij}(\vec{k}) = e \Omega_{k_j u_i}.
\end{equation}
Relating the magnetization Eq.~\eqref{eq:M_orbital} to the magnetic dipole moment of a phonon allows us to write
\begin{equation}
    M_z^{(1)} = \Gamma\,L_z,
\end{equation}
where we introduce the effective gyromagnetic ratio
\begin{equation}\label{eq:gamma}
    \Gamma = - \frac{1}{2e} \int \frac{d\vec{k}}{(2\pi)^2}\,\det \underline{Z},
\end{equation}
where $\underline{Z}$ is the Born effective charge density tensor.
The phononic gyromagnetic ratio was introduced in connection with dynamical multiferroicity~\cite{juraschek2017dynamical, Juraschek2019OrbitalPhonons, Geilhufe2021DynamicallyKTaO_3}. Eq.~\eqref{eq:gamma} constitutes a generalization of this framework in the context of ``modern theories''. $M_z^{(1)}$ takes the role of a \textit{spontaneous magnetization}, i.e., it describes the magnetization of a sample due to ionic magnetic dipoles. 

\subsection{Induced magnetization of axial phonons}
We now discuss the second term $M_z^{(2)}$ given in Eq.~\eqref{eq:M_topological} and start by connecting it to
the previously derived effective Hamiltonian Eq.~\eqref{eq:H_eff_comparison}. 
For simplicity, we focus on the adiabatic result derived in Eq.~\eqref{eq:H_L_Berry_curvature}. 
Based on that, the matrix elements for a multiplet with energy $\epsilon_0$ are 
\begin{equation}\label{eq:H_Berry_curv_4B}
    H^{ab} = \epsilon_0(\vec{k}) \delta_{ab} + \frac{\hbar}{2}L_z \Omega_{u_xu_y}^{ab}.
\end{equation}
From the definition of the Berry curvature in Eq.~\eqref{eq:BerryCurv}, we conclude that $\Omega_{u_xu_y}^{ab} = -\Omega_{u_xu_y}^{ba}$.  
For the special case of a two-fold degenerate multiplet, the resulting energy eigenvalues are
\begin{equation}
    E_\pm(\vec{k}) = \epsilon_0(\vec{k}) \pm \frac{\hbar}{2} \left|L_z \Omega_{u_xu_y}^{ab}\right|, 
\end{equation}
with the energy difference $\Delta E = \hbar \left|L_z \Omega_{u_xu_y}^{ab}\right|$. This energy difference is analogous to an orbital Zeeman effect, with the effective magnetic field,
\begin{equation}\label{eq:Beff}
    B^{\text{eff}} = \frac{\Delta E}{\mu_B} = \frac{\hbar}{\mu_B} \left|L_z \Omega_{u_xu_y}^{ab}\right|,
\end{equation}
where $\mu_B$ is the Bohr magneton, i.e., the size of the magnetic dipole moment of an electron. 
Assuming that this effective magnetic field is spatially dependent, it induces a spatially dependent magnetization via the relation
\begin{equation}
    M_z^{(2)}(\vec{r}) = \chi(\vec{r}) B^{\text{eff}}(\vec{r}).
\end{equation}
Its Fourier transform is
\begin{equation}
    M_z^{(2)}(\vec{q}) = \int \frac{d\vec{k}}{(2\pi)^2}\,\chi(\vec{q}-\vec{k}) B^{\text{eff}}(\vec{k}).
\end{equation}
Assuming homogeneous magnetization, we set $\vec{q}=0$:
\begin{equation}\label{eq:finalM2}
    M_z^{(2)} = \int \frac{d\vec{k}}{(2\pi)^2}\,\chi(-\vec{k}) B^{\text{eff}}(\vec{k}).
\end{equation}
The combination of Eqs.~\eqref{eq:M_topological} and \eqref{eq:Beff} allows us to motivate $M_z^{(2)}$ by identifying $\chi$ as the proportionality quantity between $M_z^{(2)}$ and $B^\text{eff}$, and arrive at: 
\begin{equation}
    \chi = -\frac{e^2}{4m_e} \Omega_{k_xk_y}.
\end{equation}
With this definition, the Eqs.~\eqref{eq:finalM2} and \eqref{eq:M_topological} are equivalent. As a result, we identify $M_z^{(2)}$ as the \textit{induced magnetization} due to the effective magnetic field experienced by electrons in the presence of axial phonons. 

\section{Estimate of effective magnetic field in {SrTiO}$_3$}\label{sec:example}

\begin{figure*}
    \centering
\includegraphics[width=0.95\textwidth]{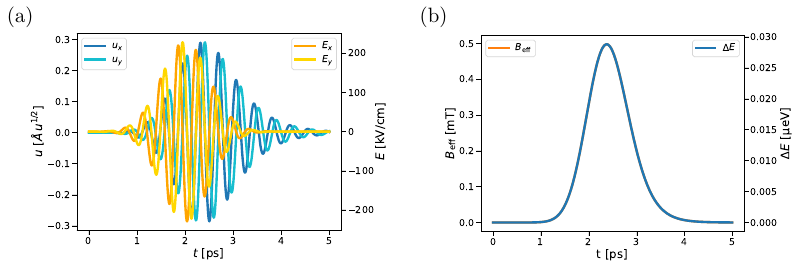}
    \caption{a) The laser drive and the corresponding phonon displacement as a function of time. The laser drive is modelled as a Gaussian pulse with a maximum amplitude of $E = \beta E_0$, with $\beta = 0.7$ and $E_0 = 230$ \si{\kilo\volt\per\centi\meter}.
    b) The energy level splitting $\Delta E$ and the corresponding effective magnetic field, $B_{\text{eff}}$, as a function of time, calculated from the phonon displacement plotted in a) with the electron-phonon coupling strength of $|g| = 10$ \si{\milli \electronvolt} and the transition energy $E_{dp}= 4.2$ \si{\electronvolt}.
    }
    \label{fig:example}
\end{figure*}
In this section, we provide an estimate of the magnitude of the effective magnetic field in $\mathrm{SrTiO_3}$, which has been measured experimentally~\cite{Basini2024}. 
To this end, we consider the level splitting of $p$-orbitals by evaluating the Hamiltonian element $H_\text{eff}^{p_x p_y}$ within the orbital basis $(s, p_x, p_y, d_{xy})$.
Oxygen $p$-orbitals are the highest-lying valence orbitals in $\mathrm{SrTiO_3}$, while titanium $d$-orbitals form the lowest-lying conduction band. For an allowed transition, the matrix element $\bra{i} \partial_{u_\alpha} U \ket{j}$ is non-zero.
This is only possible if the derivative of the potential $\partial_{u_\alpha} U$ matches the symmetry of the orbitals such that the matrix element itself becomes even under spatial inversion.
Applying this to the subspace including the $p$-orbitals, the effective Hamiltonian becomes
\begin{multline}\label{eq:H_eff_pxpy}
H_\text{eff}^{p_x p_y} = \frac{u^2}{2} i \hbar \omega \frac{\bra{p_x} \partial_{u_x} U \ket{s} \bra{s} \partial_{u_y} U \ket{p_y}}{E_{sp}^2 - \hbar^2 \omega^2} \\
- \frac{u^2}{2} i \hbar \omega \frac{\bra{p_x} \partial_{u_y} U \ket{d_{xy}} \bra{d_{xy}} \partial_{u_x} U \ket{p_y}}{E_{dp}^2 - \hbar^2 \omega^2}.
\end{multline}
The first term corresponds to the $2p \to 2s$ transition in oxygen, which would be a transition from a lower valence band to the upper valence band. However, assuming that the valence bands are completely occupied, this transition is not allowed by the Pauli exclusion principle. Therefore, we drop the first term. Furthermore, since $p_x$ and $p_y$ are degenerate, we assume that the remaining matrix elements satisfy $\bra{p_x} \partial_{u_y} U\ket{d_{xy}} = \bra{p_y} \partial_{u_x} U \ket{d_{xy}}$, therefore we can set $\bra{p_x} \partial_{u_y} U \ket{d_{xy}} \bra{d_{xy}} \partial_{u_x} U \ket{p_y} = |g|^2$.
We further assume a phonon frequency of $\omega \approx 2.7\ \mathrm{THz}$~\cite{vogt1995refined, Basini2024}, which corresponds to $\hbar \omega \approx 11\ \mathrm{meV}$. This allows us to calculate the splitting of $p_x$ and $p_y$ orbitals caused by the axial phonons as
\begin{equation}\label{eq:splitting}
    \Delta E = u^2  \frac{\hbar \omega|g|^2}{E_{dp}^2 - \hbar^2 \omega^2}.
\end{equation}
By analogy with the orbital Zeeman effect, we calculate the effective magnetic field as 
\begin{equation}\label{eq:B_eff}
    B_\text{eff} 
    = \frac{1}{\mu_B} \frac{\hbar \omega|g|^2 u^2}{E_{dp}^2 - \hbar^2 \omega^2}
\end{equation}
similar to Ref. \cite{shabala2024phonon}. 

In order to obtain an estimate for $u^2$, we solve the equations of motion for the phonon amplitude $u$, modeled as a damped harmonic oscillator
\begin{equation}\label{eq:eom}
\ddot{\vec{u}} + \eta \dot{\vec{u}} + \omega_0^2 \vec{u} = \tilde{\vec{F}},
\end{equation}
where the driving force is a Gaussian THz pulse
\begin{equation}
\tilde{\vec{F}} = \tilde{Z} \beta \vec{E} \exp\Big[-\frac{1}{2}\frac{(t-t_0)^2}{\tau^2}\Big].
\end{equation}
Here, we use an electric field strength given by $\vec{E} = E_0 (\cos{(\omega t}), \sin{(\omega t), 0})^\text{T}$, with $E_0 =  230\ \mathrm{kV/cm}$, in order to be consistent with the experiment in Ref. \cite{Basini2024}. 
We set the rest of the parameters based on the same references and assume an effective charge of $\tilde{Z} = 1.54 \,e/\sqrt{\mathrm{amu}}$, a screening constant $\beta = 0.7$, a laser pulse width $\tau = 0.5 \ \mathrm{ps}$, and a peak at $t_0 = 2$ ps.
We assume that the laser frequency is in resonance with the phonon frequency, i.e., $\omega = \omega_0 \approx 2.7\ \mathrm{THz}$ and a phonon damping constant of $\eta \approx 0.6\ \mathrm{THz}$ \cite{vogt1995refined}.
Further, we set the initial conditions as $u(0) = \dot{u}(0) = 0$. The only unknown parameter is the electron-phonon coupling strength $|g|^2$. 
For the ferroelectric soft mode in SrTiO$_3$, it is reasonable to assume that $|g|$ is on the order of $|g| \approx 10$ \si{\milli \electronvolt} \cite{Zhou2018SrTiO3}.

With these parameters, we numerically solve the equation of motion in Eq.~\eqref{eq:eom} using the Runge-Kutta method. The resulting plot of the phonon amplitude against time is presented in Fig.~\ref{fig:example}(a), where the electric field strength of the laser drive against time is also presented. 
As this figure demonstrates, the phonon amplitude reaches a peak of $u \approx 0.3$ \si{\angstrom\sqrt{\text{a.m.u.}}}.
With the obtained time evolution of $u$ it is also possible to plot the energy level splitting $\Delta E$ using Eq.~\eqref{eq:splitting}, as well as the corresponding effective magnetic field with Eq.~\eqref{eq:B_eff}.
The resulting plot is shown in Fig. \ref{fig:example}(b), where we present the effective magnetic field and the splitting as functions of time.

As can be seen in Fig.~\ref{fig:example}(b), 
the resulting effective magnetic field reaches values of $B_\text{eff} \approx 0.5$ \si{\milli \tesla}, which is two orders of magnitude smaller than the effective magnetic field reported from the experiment in Ref. \cite{Basini2024}. For the effective magnetic field to reach $\sim 30$ \si{\milli \tesla} reported in Ref. \cite{Basini2024}, the electron-phonon coupling strength would need to be approximately $80$ \si{\milli \electronvolt}.
Considering the fact that the electron-phonon coupling strength near the soft mode appears to be changing rapidly as the wave vector is varied \cite{Zhou2018SrTiO3}, a more precise estimate of this quantity could shed light on the predictions of the effective magnetic field obtained using the methods in this paper.

\section{Summary and Outlook}\label{sec:outlook}
We presented three microscopic approaches to the phonon-magnetic effect arising from the electron-phonon coupling. In the different approaches, the coupling was treated as a weak perturbation, as an adiabatic change to the Hamiltonian, and as a time-periodic term. Some of these derivations were given in more specialized cases in the recent literature~\cite{shabala2024phonon, klebl2024ultrafast, ren2021phonon}, which are now put on an equal footing. In our derivation, the Floquet and the perturbation methods give the same final effective Hamiltonian, while the adiabatic method yields the same result in the low-frequency limit. This consistency shows that these theories are in agreement with each other and that studying the role of the electron-phonon interaction is a promising direction in understanding the phono-magnetic effect.

We further showed the connection between the phonomagnetic moment, the effective magnetic field, and the sample magnetization. Calculating the expected effective magnetic field for the soft mode in SrTiO$_3$ resulted in a prediction that was two orders of magnitude lower than the experimental observations \cite{Basini2024}, for an electron-phonon coupling strength in agreement with the computations of Ref.~\cite{Zhou2018SrTiO3}. However, this could be caused by underestimating the size of the electron-phonon coupling strength $|g|$. Therefore, in order to evaluate how reliable these theories are in predicting the size of the phono-magnetic effects, a precise value of the electron-phonon matrix elements is needed, which, hopefully, inspires future work. 

\begin{acknowledgements}
We acknowledge support from the the Knut and Alice Wallenberg Foundation (Grant No. 2023.0087), the Swedish Research Council (VR starting Grant No. 2022-03350), the Olle Engkvist Foundation (Grant No. 229-0443), as well as the department of physics and the areas of advance Nano and Material Science at Chalmers University of Technology. 
\end{acknowledgements}

\end{document}